# Self-Configuring Universal Linear Optical Component

**David A. B. Miller**

*Ginzton Laboratory, Stanford University, 348 Via Pueblo Mall, Stanford CA 94305-4088*
*Corresponding author: dabm@ee.stanford.edu*

We show how to design an optical device that can perform any linear function or coupling between inputs and outputs. This design method is progressive, requiring no global optimization. We also show how the device can configure itself progressively, avoiding design calculations and allowing the device to stabilize itself against drifts in component properties and to continually adjust itself to changing conditions. This self-configuration operates by training with the desired pairs of orthogonal input and output functions, using sets of detectors and local feedback loops to set individual optical elements within the device, with no global feedback or multiparameter optimization required. Simple mappings, such as spatial mode conversions and polarization control, can be implemented using standard planar integrated optics. In the spirit of a universal machine, we show that other linear operations, including frequency and time mappings, as well as non-reciprocal operation, are possible in principle, even if very challenging in practice, thus proving there is at least one constructive design for any conceivable linear optical component; such a universal device can also be self-configuring. This approach is general for linear waves, and could be applied to microwaves, acoustics and quantum mechanical superpositions.

## 1. INTRODUCTION

There has been growing recent interest in optical devices that can perform novel functions such as converting spatial modes from one form to another [1-3], offering new kinds of optical frequency filtering [4-7], providing optical delays [8,9], or enabling invisibility cloaking [10-13]. All these operations are linear. Many other linear transformations on waves are mathematically conceivable, involving frequency or time, spatial form, polarization, and non-reciprocal operations. Despite the mathematical simplicity of defining such linear operations [14], it has not been clear how to perform arbitrary linear operations on waves physically, or even in principle whether such operations are generally possible. The usual linear optical components, such as lenses, mirrors, gratings, and filters, only implement a subset of all the possible linear relations between inputs and outputs [15]. Other components such as volume holograms [16, 17] or matrix-vector multipliers [18] can implement some more complex relations; it is difficult, however, to make such approaches efficient – for example, avoiding a loss factor of $1/M$ when working with $M$ different beams [3]; for high-efficiency devices, interactions between designs for different inputs leave it unclear how, or even if, the device is possible. Indeed, some designs resort to blind optimization based in part on random or exhaustive searches among designs with no guarantee of the existence of any solution [4-7, 19, 20]; such approaches do, however, give some existence proofs of the possibility of some efficient designs for novel functions [4, 5, 7, 19, 20].

In this paper, we show how to design an arbitrary linear optical device. The method is direct and progressive; once we decide what we want the device to do, we sequentially set the various required components one by one. For devices operating only on spatial modes, the devices could be made using standard optics and are particularly well-suited to integrated optical approaches. In this spatial case, we can describe the device as a general spatial mode converter. The spatial approach can be extended to handle polarization by converting different polarizations in the same spatial input mode to the same polarization in two different spatial modes and proceeding thereafter in a similar fashion to the spatial mode converters.

More generally for a linear optical device, we have to look beyond fixed spatial structures to ones that also vary in time. Note, for example, that a device with a refractive index that varies in a prescribed way in time can be linear in the signal field in mathematically the same way as a device where that index only varies in space. Just as a linear spatial optical device can map an input beam with one shape to an output beam with another shape, so also in principle can a linear temporal (i.e., time-varying) optical device map an input beam with one spectrum to an output beam with another spectrum. Such temporal devices would require frequency shifters or time-slot interchangers or equivalent temporal operations. With current technologies, it is practically much more difficult to make the required large changes in optical properties at timescales corresponding to optical frequencies, of course, As a result, most conceivable temporal linear optical devices are not currently practical. It is, however, of some basic interest to understand at least what devcies are possible in principle. Here, in an extension of the discussion of spatial and polarization devices, we also examine such temporal devices. We show a constructive design method that could in principle design any linear optical device, including temporal aspects, without violating any laws of physics. In the spirit of a universal machine, like the Turing machine in computing, we therefore prove here that any such device is possible in principle. , Even with some design approach for an arbitrary desired linear operation, the resulting device could be quite complicated [15] and the design could require a significant amount of calculation. Furthermore, operations on waves can require interferometric precision, and calibrating and setting many analog elements precisely to construct such a design could be very challenging even for the spatial mode converter devices.

Fortunately, we can avoid the calculations and the difficulties of calibration and setting of devices; we can make the device self-configuring. The self-configuration involves training the device using the desired inputs and output. It is based on an extension of the ideas of the self-aligned beam coupler [21]. This process requires only local feedback loops each operating on a single measureable parameter. Such feedback loops can be left running during device operation, allowing continuous optimization and compensation for drifts in devices. This self-configuration is also progressive, requiring no global calculations or optimization. This self-configuration also applies in principle to the temporal devices. Note that, as a result of this self-configuration, arbitrary linear optical devices can be designed without performing any calculations. Instead, we need only simple progressive training operations.

In this paper, in Section 2, we first describe the general spatial mode converter and its extensions to handling polarizations. We describe this using the self-configuring approach; this actually involves less mathematics than a direct calculation of the required design, which we defer to Appendix A. (Detailed analysis of Mach-Zehnder interferometers for use in the approach is given in Appendix B.) In Section 3, we discuss the underlying linear algebra of the approach, showing how it relates directly to the general description of linear optical devices [14] and to the related counting of complexity [15]. The generalization of the device to handle wavelength or frequency attributes is discussed in Section 4, including self-configuring operation in these cases also. (An alternative time- rather than frequency-based approach is given in Appendix C, and non-reciprocal devices are discussed in Appendix D.) We draw conclusions in Section 5.

## 2. Device concept for spatial beams

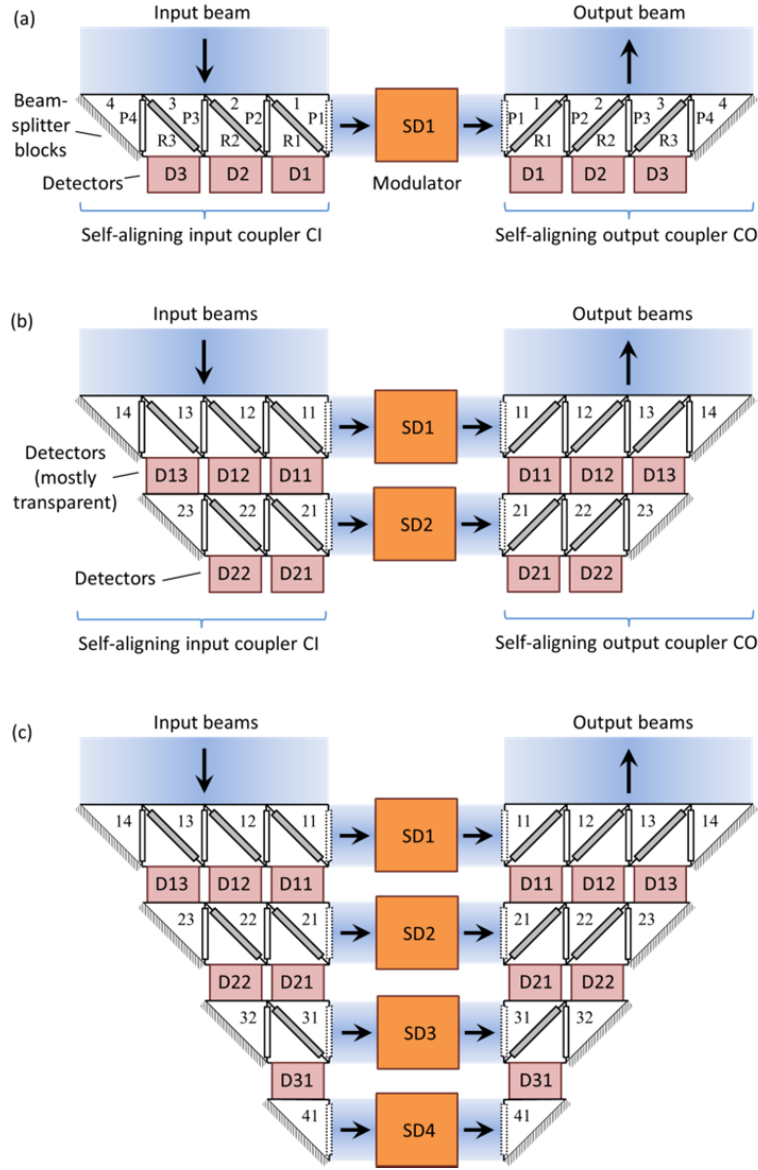

Fig. 1. Schematic illustration of the self-configuring device. Diagonal grey rectangles are controllable reflectors. Vertical clear rectangles are controllable phase shifters. Dashed clear rectangles are optional phase shifters that may be present in the implementation, but are not necessary. Configurations for (a) one input and output beam pair, (b) two beam pairs, (c) all four possible beam pairs.

The concept of the approach for an arbitrary device operating on spatial modes (a general spatial mode converter) is shown in Fig. 1, illustrated here first for an example with the inputs and outputs sampled to four channels. It consists of two self-aligning universal beam couplers [21], one, CI, at the input, and another, CO, at the output. These are connected back-to-back through modulators that can set amplitude and phase; these modulators could also incorporate gain elements. The self-aligning couplers require controllable reflectors and phase shifters together with photodetectors that are connected in selectable feedback loops to control the reflectors and phase shifters [21]. (Dashed rectangle phase shifters are not required, but may be present depending on the way the devices are implemented, and might be desirable for symmetry and equality of path lengths.)

We presume that, for our optical device, we know what set of orthogonal inputs we want to connect, one by one, to what set of orthogonal outputs. If we know what we want the component to do, any linear component can be completely described this way, as discussed in [14]. The simplest case is that we want the device to convert from one specific spatial input mode to one specific spatial output mode (Fig. 1(a)).

## 2.1 Single beam case

To train the device as in Fig. 1(a), we first shine the specific input mode or beam of interest onto the top of the input self-aligning coupler CI. Then we proceed to set the phases and reflectivities in the beam splitter blocks in CI as in [21]. Briefly, this involves first setting phase shifter P4 to minimize the power in detector D3; this aligns the relative phases of the transmitted and reflected beams from the bottom of beamsplitter 3 so that they are opposite, therefore giving maximum destructive interference. Then we set the reflectivity R3 to minimize the D3 signal again; presuming that the change of reflectivity makes no change in phase, the D3 signal will now be zero because of complete cancellation of the reflected and transmitted light shining into it. Next, we set phase shifter P3 to minimize the D2 signal, then adjust R2 to minimize the D2 signal again. Proceeding along all the beamsplitter blocks in this way will lead to all the power in the input mode emerging in the single output beam on the right.

The second part of the training is to shine a reversed (technically, phase-conjugated [24-26]) version of the desired specific output mode onto the output coupler CO; that is, if we want some specific mode to emerge from the device (i.e., out of the top of CO), then we should at this point shine that mode back into this "output". We set the values of the phase shifters and reflectivities in coupler CO by a similar process to that used for coupler CI, which will lead to this "reversed" beam emerging from the left of the row of beamsplitter blocks, for the moment going backwards into modulator SD1 from the right.

Now that we have set reflectivity and phase values in coupler CO, to understand what we have accomplished for coupler CO, we imagine that we turn off the training beam that was shining backwards onto the top of coupler CO and shine a simple beam instead from the output of modulator SD1 into CO. It is obvious that will lead to all the power coming out of the top of coupler CO and that the resulting field magnitudes (and powers) of beams emitted from the tops of the beamsplitters will be the same as the ones incident during the training. To understand why the phases are set using a phase conjugate beam during training, we can formally derive the mathematics of the design, as discussed in Appendix A; we can, however, also understand this intuitively. Suppose, for example, that during training the (backward) beam incident on the top of beamsplitter block 4 (of CO) had a slight relative phase lead compared to that incident on the top of beamsplitter block 3 (as would be the case if it was a plane wave incident from the top right). Then, during training, we would have added a relative phase delay in phase shifter P4 to achieved constructive interference of these two different input portions as we move along the line of beam splitters. Running instead in the "forward" mode of operation, then, the beam that emerges vertically from beamsplitter block 4 will now have a phase delay compared to that emerging from block 3 (as would be the case if it was a plane wave heading out to the top right). The resulting phase front emerging from the top of coupler CO is therefore of the same shape (at least in this sampled version) as the backward (phase conjugated) beam we used in training, but propagating in the opposite direction as desired.

So, with the device trained in this way, shining the desired input mode onto CI will lead to the desired output mode emerging from CO. Finally, we set modulator SD1 to get the desired overall amplitude and phase in the emerging beam; choosing these is the only part of this process that does not set itself during the training. Modulator SD1 could also be used to impose a modulation on the output beam, and an amplifier could also be incorporated here if desired for larger output power.

### 2.2 Operating with multiple beams simultaneously

The process can be extended to more than one orthogonal beam. For beams under conventional optical conditions (e.g., avoiding near-field effects), orthogonality can usually be sufficiently understood in terms of the orthogonality of the electric field patterns of the modes (for each polarization, if necessary). Our descriptions below take this approach. More generally, we can always unambiguously establish orthogonal modes for a device by evaluating the communications modes of the coupling operators from the original beam source and to the final wave receiving volume, as discussed in Appendix A of [15].

In Fig. 1(b), having trained the device for the desired "first" input and output beams, we can now train it similarly with a "second" pair of input and output beams that are orthogonal to the "first" beams. Since the device is now set so that all of the "first" beam shone onto the top of CI will emerge into modulator SD1, then any "second" beam that is orthogonal to the "first" beam will instead pass entirely into the photodetectors D11 – D13 (or, actually, through them, since now we make them mostly transparent, as discussed in [21]). Though this second beam is changed by passing through the top (first) row of beamsplitters, it is entirely transmitted through them to the second row of beamsplitter blocks. (Note that, provided the mostly transparent detectors D11 – D13 have substantially equal loss, that loss does not affect the orthogonality of the beam passed to the second row; such equal loss could also be compensated by introduction of gain in the modulator SD2.) In the second row of beamsplitter blocks, we can run an exactly similar alignment procedure, now using detectors D21 – D22 to minimize the signal based on adjustments of the phase shifters and reflectivities in the second row of beamsplitter blocks in coupler CI.

We can proceed similarly by shining the reversed (phase conjugated) version of the desired second (orthogonal) output beam into the top of coupler CO, running the self-alignment process similarly for this second row. Then, shining the second input beam into CI will lead to the desired second output beam emerging form CO. If our device requires us to specify more than two mode couplings, we can continue this process, adding more rows until the number of rows equals the number of blocks (here 4) in the first row. Fig. 1 (c) illustrates a device for 4 beams. Note that once we have set the device for the first 3 desired orthogonal pairs, then the final (here, fourth) orthogonal pair is automatically defined for us, as required by orthogonality. Formally, in the notation of [15], the number of rows we require in our device here has to equal the mode coupling number, $M_C$.

### 2.3 Implementation with Mach-Zehnder interferometers

Similarly to such bulk beamsplitter versions discussed in [21], the configurations in Fig. 1 are idealized. We are neglecting any diffraction inside the apparatus, we are presuming that our reflectors and phase shifters are operating equally on the entire beam segment incident on their surfaces, and we are presuming that each such beam segment is approximately uniform over the beamsplitter width. The path lengths through the structure are also not equal for all the different beam paths, which would make this device very sensitive to wavelength; different wavelengths would have different phase delays through the apparatus, so the phase shifters would have to be reset even for small changes in wavelength.

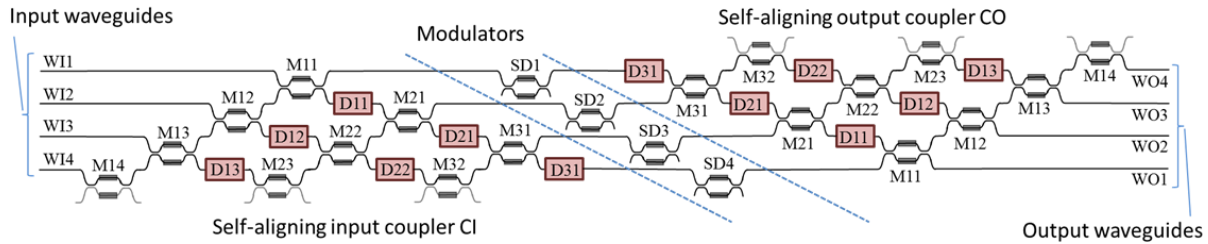

Fig. 2. Example planar layout of a device analogous to Fig. 1(c) with Mach-Zehnder interferometers providing the variable reflectivities and the phase shifts. Not shown are devices such as grating couplers that would couple different segments of the input and output beams into and out of the waveguides WI1 – WI4 and WO1 – WO4, respectively. The self-aligning output coupler CO is reflected about a horizontal axis compared to Fig. 1(c) for compactness. Greyed arms of Mach-Zehnder interferometers M14, M23, and M32 in both the input (CI) and the output (CO) self-aligning couplers are optional; these devices are operated only as phase shifters and could be replaced by simple phase shifters.

An alternative and more practical solution is to use Mach-Zehnder interferometers (MZIs) in a waveguide configuration [21]; diffraction inside the apparatus is then avoided, and equalizing waveguide lengths can eliminate the excessive sensitivity to wavelength. Fig. 2 illustrates such a planar optics configuration. Common mode (i.e., equal) drive of the two phase shifting arms of such an MZI changes the phase of the output; differential (i.e., opposite) drive of the arms changes the "reflectivity" (i.e., the split ratio between the outputs) (see Appendix B for a detailed discussion of the properties of the MZIs as phase shifters and variable reflectors).

The configuration in Fig. 2 formally differs mathematically from that in Fig. 1(c) in that we have reflected the output self-aligning coupler CO about a horizontal axis to achieve a more compact device. This reflection makes no difference to the operation of the device; since the device can couple arbitrary beams, the labeling or ordering of the waveguides is of no importance. Note only that the order of the output beams is reversed compared to the input beams – this device links input beam 1 to output beam 4, and so on. (This reflection would be equivalent to similarly reflecting the self-aligning output coupler CO in Fig. 1 about a horizontal axis, which would lead to the output beam coming out of the bottom, rather than the top, of the device.) Schemes are also discussed in [21] for ensuring equal numbers of MZIs in all optical paths for greater path length and loss equality by the insertion of dummy devices, and such schemes could be implemented here also.

The use of sets of grating couplers connected to the input waveguides WI1 − WI4 and to the output waveguides WO1 − WO4 is one way in which this device could be connected to the input and output beams, as discussed in [21]. In this case, though the wave is still sampled at only a finite number of points or regions, we can at least obtain true cancellation of the fields in the single mode guides even if the field on the grating couplers is not actually uniform. The geometry of Fig. 2 also shows that we can make a device that has substantially equal time delays between all inputs and outputs because all the waveguide paths are essentially the same length. As discussed in [21], such equality is important if the device is to operate over a broad bandwidth.

The example so far has considered a beam varying only in the horizontal direction, and using only four segments to represent the beam. Of course, the number of segments we need to use depends on the complexity of the linear device we want to make [15], and the number could well be much larger than 4; we will discuss such complexities in Section 3. Additionally, we would likely want to be able to work with two-dimensional beams, in which case we could imagine two-dimensional arrays of grating couplers coupling into the one-dimensional arrays of waveguides of Fig. 2, as discussed in [21].

### 2.4 Extension to polarization

So far, we have only discussed linear devices operating on spatial modes. We can relatively simply extend the concept to include polarization as well. Consider a polarization converter as in Fig. 3. In this example, an incident beam of the desired polarization is split into two orthogonal polarizations, for example, using a polarization demultiplexing grating coupler [22]. The polarization demultiplexer here is converting the physical representation from a polarization basis on a single spatial mode to a representation in two spatial modes (the waveguide modes) on a single polarization. Then the simple two-channel self-aligning coupler CI can combine the fields and powers from the two polarizations in this one particular beam loss-lessly into one single-mode waveguide WIO. Here, as before, we adjust the phase shifter PI to minimize the power in detector DI, and then adjust the "reflectivity" of the Mach-Zehnder interferometer MZI (by differential drive of the phase shifters in the two arms) to minimize the power in detector DI.

In many situations, this may be the desired output, and we could take this output from waveguide WIO at the point of the dashed line in Fig. 3. We could therefore run this device as a polarization stabilizer; leaving the feedback sequence running continuously, the output will remain in the single polarization in the waveguide WIO even if the input polarization state drifts. Note that, in contrast to common polarization state controllers (e.g., [23]), this device requires no global feedback loop and no simultaneous multiple parameter optimization. It also requires no calculation [23] in the feedback loop.

If we wish, instead, we can change the wave from the output grating coupler into any desired polarization using the second, output self-aligned coupler CO; we can program this desired output polarization by training with the desired polarization state running backwards into that output grating coupler and running the feedback loops with PO, MZO and DO in the same way as we did for the input. With this device operating with circular polarizations, if we train with a right-circular polarization going backwards "in" to the output coupler from the outside, for example, the action of the device under "forwards" operation is such that the beam emerging from the output coupler will also be right circularly polarized. (Note that, if we only want this polarization conversion, DI and DO could be the same detector.)

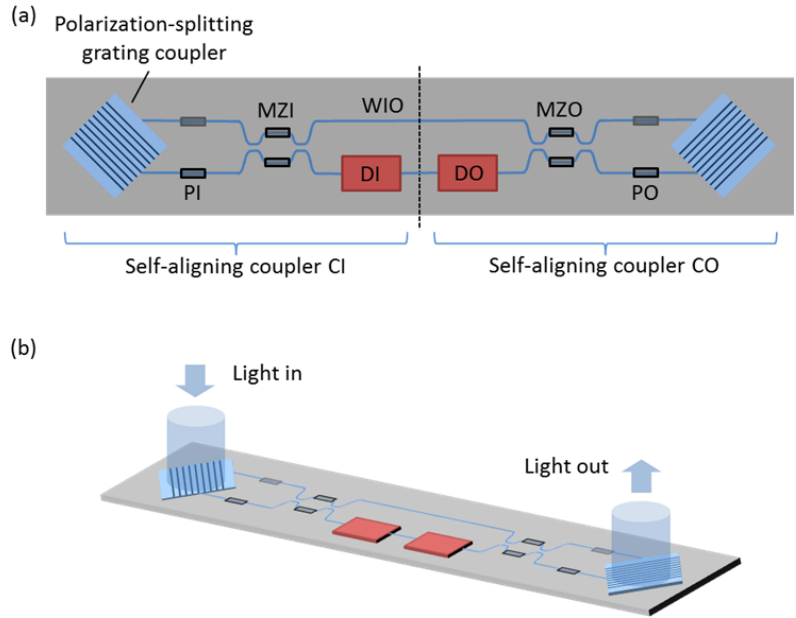

Fig. 3. Polarization converter. (a) Plan view. (b) Perspective view. Light incident on the grating coupler in self-aligning coupler CI is split by its incident polarization into the two waveguides, and similarly light from the waveguides going into the grating coupler in self-aligning coupler CO appears on the two different polarizations on the output light beam. PI and PO are phase shifters; the similar but greyed boxes are optional dummy phase shifters. Optionally, a phase shifter and its dummy partner could instead be driven in push-pull to double the available relative phase shift. MZI and MZO are Mach-Zehnder interferometers, and DI and DO are detectors.

For example, a right circularly polarized beam can be considered to have the vertical linear polarization leading the (right-pointing) horizontal polarization by 90 degrees. (Note that when we say "right-pointing" we mean relative to the direction of propagation.) So, when shining a training beam backwards into the coupler CO we would impose a 90 degree phase delay on the vertical polarization when aligning the coupler to achieve constructive interference as we combine them into the desired (backwards) linearly polarized mode in WIO in coupler CO. Running a beam now forwards in waveguide WIO into CO will mean that the emerging beam has the vertical polarization lagging by 90 degrees compared to the horizontal polarization because of our introduced phase delay; however, that horizontal polarization is now left-pointing relative to the (now forwards) direction of propagation, which means the vertical polarization leads a right-pointing horizontal polarization by 90 degrees, which is right circular polarization again in the outgoing beam. This is analogous to the behavior of a phase-conjugating mirror; in contrast, a conventional mirror would change right-circular polarization to left-circular on reflection.

If we make the detectors DI and DO mostly transparent and join them with a waveguide as shown to allow transmission through them from left to right, then this device converts from one set of orthogonal polarization states at the input to another set of orthogonal polarization states at the output. For example, if we had trained the device to convert from right-circularly polarized light at the input to vertical linear polarization at the output, then left-circularly polarized light at the input would appear as horizontal polarization at the output.

We could also choose to add modulators in the waveguide WIO and the waveguide between the photodetectors in Fig. 3, which would allow us to make a fully arbitrary polarization device. In this case, the device would be selecting two orthogonal polarization channels that we could choose arbitrarily, allowing separate modulation of these two channels, and presenting them at the output as two orthogonal polarization channels of our choice. (Again, we could operate this system with DI and DO as the same mostly-transparent detector.)

## 3. Mathematical discussion

At this point, we can usefully relate this device explicitly to the mathematical description of linear devices in [14] and the counting of complexity in [15].

### 3.1 General mathematics of linear optical devices

Quite generally [14, 15], any linear optical device can be described mathematically in terms of a linear "device" operator $D$ that relates an input wave, $|\phi_I\rangle$, to an output wave $|\phi_O\rangle$ through

$$|\phi_O\rangle = D|\phi_I\rangle \tag{1}$$

It can be shown [14] that essentially any such linear operator $D$ corresponding to a linear physical wave interaction in a device can be factorized using the singular value decomposition (SVD) to yield an expression

$$D = \sum_m s_{Dm} |\phi_{DOm}\rangle \langle \phi_{DIm}| \tag{2}$$

or, equivalently,

$$D = V D_{diag} U^\dagger \tag{3}$$

Here $U$ is a unitary operator that in matrix form has the vectors $|\phi_{DIm}\rangle$ as its column vectors, and similarly $|\phi_{DOm}\rangle$ are the column vectors of the matrix for the unitary operator $V$. $D_{diag}$ is a diagonal matrix with complex elements (the singular values) $s_{Dm}$. The sets of vectors $|\phi_{DIm}\rangle$ and $|\phi_{DOm}\rangle$ form complete orthonormal sets for describing the input and output mathematical spaces $H_I$ and $H_O$ respectively [14] (at least if we restrict those spaces to containing only those functions that can be connected using the device).

The resulting singular values are uniquely specified, and the unitary operators $U$ and $V$ (and hence the sets $|\phi_{DIm}\rangle$ and $|\phi_{DOm}\rangle$) are also unique (at least within phase factors and orthogonal linear combinations of functions corresponding to the same magnitude of singular value, as is usual in degenerate eigenvalue problems). An input $|\phi_{DIm}\rangle$ leads to an output $s_{Dm}|\phi_{DOm}\rangle$, so these pairs of vectors define the orthogonal (mode-converter) [14] "channels" through the device.

In a practical device, we may have a physical input space that we would describe with $M_I$ modes or basis functions and similarly an output space that we would describe using $M_O$ modes or basis functions. For example, the input mathematical space might consist of a set of $M_I$ Gauss-Laguerre angular momentum beams, and the output space might be a set of $M_O$ waveguide modes or $M_O$ different single-mode waveguides, with $M_I$ and $M_O$ not necessarily the same number. As another example, we might be describing the input space with a set of $M_I$ waveguide modes, and the output space might be described with a plane-wave or Fourier basis of $M_O$ functions, as appropriate for free-space propagation. In any of these cases, the actual number of orthogonal channels, $M_C$, going through the device (the "mode coupling number" $M_C$ in the notation of [15] might be smaller than either $M_I$ or $M_O$ (or both) . For example, we could have large plane wave basis sets for describing the input and output fields of a 3-moded waveguide; no matter how big these input and output sets are, however, there will only practically be $M_C = 3$ orthogonal channels through the device. In the notation of [15], if $M_C$ is equal to the smaller of $M_I$ or $M_O$, then the device is "maximally connected" – it has the largest number of possible orthogonal channels from input to output given the dimensionalities of the input and output spaces.

### 3.2 Mathematics of general spatial mode converters

In the example devices of Fig. 1, the most obvious choices for the input and output basis function sets are the "rectangular" functions that correspond to uniform waves that fill exactly the (top) surface of each single beamsplitter block; in this example, we have chosen equal numbers ($M_I$ and $M_O$ each equal to 4) of such blocks on both the input and the output, though there is no general requirement to do that, and the number $M_C$ of channels through the device is the number of rows of beamsplitter blocks (1 in Fig. 1(a), 2 in Fig. 1(b), and 4 in Fig. 1(c)). In those devices also, the (complex) transmissions of the modulators SD1 – SD4 correspond mathematically to the singular values $s_{Dm}$.

In these cases of possibly different values for each of $M_I$, $M_O$, and $M_C$ it is more useful and meaningful to define the matrix $\mathsf{U}$ as an $M_I \times M_C$ matrix (so $\mathsf{U}^\dagger$ is a $M_C \times M_I$ matrix)  and the matrix $\mathsf{V}$ as an $M_O \times M_C$ matrix. With these choices, the matrix $\mathsf{D}_{diag}$ becomes the $M_C \times M_C$ square diagonal matrix with the (generally non-zero) singular values $s_{Dm}$ as its elements. If there are only $M_C$ possible orthogonal channels through the device, then there are only $M_C$ singular values that are possibly non-zero also. As discussed in [15], using these possibly rectangular (rather than square) forms for $\mathsf{U}$ and/or $\mathsf{V}$ means we are only working with the channels that could potentially have non-zero couplings (of strengths given by the singular values) between inputs and outputs.

In the device of Fig. 1, the input coupler CI corresponds to the matrix $\mathsf{U}^\dagger$, the vertical line of modulators corresponds to the diagonal line of possibly non-zero diagonal elements in $\mathsf{D}_{diag}$, and the output coupler CO corresponds to the matrix $\mathsf{V}$. In the cases of Figs. 1(a) and 1(b), the matrices $\mathsf{U}$ and $\mathsf{V}$ are not square. Because they are not square, in this amended way of writing the mathematics, they are not therefore unitary, as discussed in [15]. We have, however, we eliminated elements in our mathematics that serve no purpose; we have essentially avoided having our mathematics describe rows of beam splitters and modulators that do not exist physically. For example, for a two channel (i.e., two beam) device as in Fig. 1(b), we could write the form as in Eq. (3) as

$$D = \begin{bmatrix} v_{11} & v_{12} \\ v_{21} & v_{22} \\ v_{31} & v_{32} \\ v_{41} & v_{42} \end{bmatrix} \begin{bmatrix} s_{D1} & 0 \\ 0 & s_{D2} \end{bmatrix} \begin{bmatrix} u_{11}^* & u_{21}^* & u_{31}^* & u_{41}^* \\ u_{12}^* & u_{22}^* & u_{32}^* & u_{42}^* \end{bmatrix} \tag{4}$$

where

$$\left| \phi_{DI1} \right\rangle = \begin{bmatrix} u_{11} \\ u_{21} \\ u_{31} \\ u_{41} \end{bmatrix} \quad \left| \phi_{DI2} \right\rangle = \begin{bmatrix} u_{12} \\ u_{22} \\ u_{32} \\ u_{42} \end{bmatrix} \quad \left| \phi_{DO1} \right\rangle = \begin{bmatrix} v_{11} \\ v_{21} \\ v_{31} \\ v_{41} \end{bmatrix} \quad \left| \phi_{DO2} \right\rangle = \begin{bmatrix} v_{12} \\ v_{22} \\ v_{32} \\ v_{42} \end{bmatrix}$$

$$\tag{5}$$

Despite that fact that $U$ and $V$ are no longer necessarily unitary, the forms of Eqs. (1) – (3) remain valid. The sets of functions $\left| \phi_{DIm} \right\rangle$ and $\left| \phi_{DOm} \right\rangle$ are complete for representing input and output functions corresponding to non-zero couplings (i.e., non-zero singular values) through this device and are still the columns of the matrices $U$ and $V$, respectively. (The settings of the phase shifters and reflectors in the full unitary forms of couplers CI and CO as shown in Fig. 1(c) would each correspond to a Gaussian-elimination-like factorization of a unitary matrix [27, 28] as discussed in [28]; other forms, such as the multilayer binary tree form in [21], would correspond to other possible factorizations of such unitary matrices.)

At this point, we can make a direct relation between the number of adjustable parameters in the physical devices in Figs. 1 and 2 and the "complexity number" $N_D$ of real numbers required to specify the device according to [15]. The number of independent real numbers required to specify the $M_I$ dimensional vector $\left| \phi_{DI1} \right\rangle$ (i.e., to choose an arbitrary specific first input beam) is $2M_I$ - 2; the "-2" is because (i) the vector is normalized, removing one degree of freedom, and (ii) the overall phase of such a vector (i.e., of the beam) is arbitrary. Note that this number corresponds exactly to the number of adjustable parameters in the devices in the first row of the self-aligning input coupler CI; in Fig. 1, $M_I = 4$, and we have 3 adjustable reflectors and 3 phase shifters, for a total of $2M_I$ - 2 = 6.

The number of independent real numbers required to specify $\left| \phi_{DI2} \right\rangle$ is smaller by 2 because $\left| \phi_{DI2} \right\rangle$ has to be orthogonal to $\left| \phi_{DI1} \right\rangle$ – i.e., both the real and imaginary parts of the inner product $\left\langle \phi_{DI2} \middle| \phi_{DI1} \right\rangle$ have to be zero – so we need $2M_I - 4$ real numbers to specify this second vector, which corresponds to the $2M_I$ - 4 = 4 adjustable elements (2 reflectors and 2 phase shifters) in the second row in Fig. 1(b) or 1(c). As discussed in [15], following this approach to counting device complexity, the total "complexity number" $N_D$ of real numbers required to specify a "maximally functional" device (i.e., one for which we can make arbitrary choices of sets of orthogonal input and output functions within the dimensionalities of the spaces) is generally

$$N_D = 2M_C \left( M_I + M_O - M_C \right) \tag{6}$$

which corresponds to the total number of physically adjustable parameters in the devices of Fig. 1. (Note there are two adjustable parameters associated with each modulator – amplitude and phase.) $M_C$ is 1 in Fig. 1(a), 2 in Fig. 1(b) and 4 in Fig. 1(c).

Though here we will emphasize the self-configuring approach, the specific settings of the phase shifters and reflectors can instead be calculated straightforwardly given the desired function of the device. See Appendix A for an explicit sequential row-by-row and block-by-block physical design process for the partial reflector and phase shifter parameters. Appendix B gives the formal analysis for the MZI implementation of variable reflectors and phase shifters.

### 3.3 Use of finite-dimensional Hilbert spaces

One final formal issue for an arbitrary device is that the input and output Hilbert function spaces, $H_I$ and $H_O$ respectively, in which $|\phi_I\rangle$ and $|\phi_O\rangle$ exist mathematically, may well each have infinite numbers of dimensions, whereas our device has finite dimensionality. To resolve this apparent discrepancy, note first that the input waves $|\phi_I\rangle$ come from some wave source in another volume (generally, a "transmitting" Hilbert space $H_T$), through some coupling operator $G_{TI}$. Because of a sum rule [14, 29], there is only a finite number of channels between $H_T$ and $H_I$ that are strongly enough coupled to be of interest. A familiar example is the practically finite number of distinct "spots" that can be formed on one surface from sources on another, consistent with diffraction [29]. A similar argument holds at the output with output waves $|\phi_O\rangle$ leading to resulting waves in some "receiving" space $H_R$. This point is discussed in greater depth in [15]. Hence, we can practically presume that $D$ can be written as a sufficiently large but finite-dimensional matrix to any degree of approximation we wish.

## 4. Universal linear device

So far, we have only considered spatial and polarization input and output modes for the device concept, though the underlying mathematics discussed above and in Refs. [14] and [15] can treat any additional linear attributes also, such as frequency or time, and could in principle also handle attributes like quantum mechanical spin in other wave systems. We can at least conceive of a universal machine that would attempt to perform any linear mapping between inputs and outputs. Mathematically, it is straightforward to construct the necessary Hilbert spaces, which would be formed by direct products of the different basis functions corresponding to each attribute separately (see, e.g., [30]).

### 4.1 Universal device with representation converters

One general approach that would work in principle for a universal device is to physically convert each such direct product basis function (e.g., one with specific spatial, temporal and polarization characteristics) to a monochromatic spatial mode, a mode we can then feed through a version of the spatial device we discussed above. In other words, we can propose that we could convert the representation to a simple monochromatic spatial one (e.g., in fiber or waveguide modes), perform the desired mathematical device operation (i.e., the mathematical operator $D$), using our spatial approach discussed above, and then convert the representation back to its full spatial, temporal and polarization form. Performing this representation conversion means that aspects of a light beam that do not normally "interfere" with one another, such as different polarizations and frequencies, can now mathematically be scattered into one another arbitrarily, as required for the most general mathematical linear operation on the optical field.

Therefore, we need to make "representation converters" to change into and back out of the single-frequency, single-polarization, waveguide mode representation we use inside our universal spatial device, or general spatial mode converter, as discussed above. The mathematical operator D that describes that mapping from input modes to output modes is not changed, but the physical representation of those modes is changed inside the device, and is changed back before we leave the device. The polarization converter discussed above employs a simple example of such a representation conversion, changing one spatial mode with two different polarizations into two spatial modes in the same polarization so that we can arbitrarily "interfere" the two polarizations inside the device.

### Example temporal device interfering two colors

Before proceeding to discussing a hypothetical fully universal linear device, because such devices can be quite unlike more common practical optical devices, it may be useful to consider a simple conceptual example. Suppose that, instead of beams of two orthogonal polarizations in one incident spatial mode as in the polarization device of Fig. 3, we have beams of two different colors – "red" and "blue" – in the same spatial mode and we make a "red-blue" interference device as shown conceptually in Fig. 4. We use "red" and "blue" figuratively here to mean two different frequencies of input light, not necessarily actual red and blue colors, though we do presume these are each monochromatic light fields.

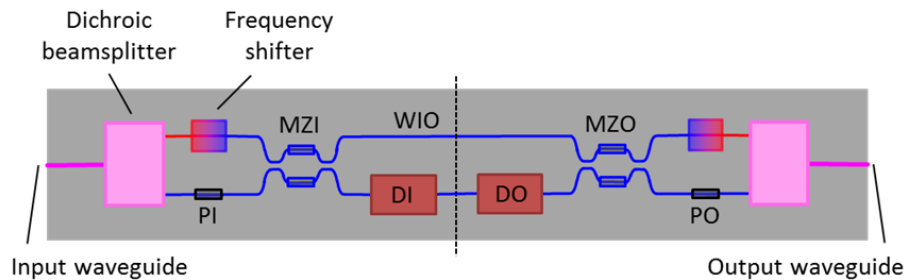

Fig. 4. Red-blue interference device. A mixture of "red" and "blue" light at the input is split into its "red" and "blue" components by a dichroic beamsplitter. Then the "red" component is converted to "blue" by a frequency shifter so both components are represented by "blue" light but in different waveguides. The device can be trained to look for any particular combination of "red" and "blue" and to output any particular combination of "red" and "blue" as a result.

Instead of a grating coupler that separates the two polarizations to two different spatial waveguides, imagine that we use a dichroic beamsplitter to separate the two colors to different waveguides. Now presume additionally that in the resulting "red" waveguide, we insert a frequency shifter that turns the "red" beam into a "blue" one – that is, it shifts the frequency of the "red" beam to be exactly that of the "blue" beam. Such frequency shifters are possible in principle [37-40] though quite challenging in practice. One conceptual approach would use a modulator arrangement driven at the difference frequency of the "red" and "blue" beams, with the modulator drive being derived from the beat signal between the original "red" and "blue" sources. We make a complementary combination of a frequency shifter and dichroic beamsplitter at the output.

Just as the polarization device in Fig. 3 can be set up to look for any particular combination of the two input polarizations and to output any particular combination of the two polarizations at the output, this device performs an analogous operation but on two colors rather than two polarizations.

For example, we could train the input side of device to look for an input that corresponded to a "red" and a "blue" beam with equal amplitude and a specific phase of their beating (as defined relative to the phase of the drive to the frequency shifter). Then, we could set the Mach-Zehnder MZO so that its output was all in the lower waveguide (i.e., through phase shifter PO) and so the output waveguide would contain only a "blue" beam; this could be accomplished by a training process in which we shine only a "blue" beam backwards into the output waveguide. (We presume here that a frequency shifter that shifts "blue" to "red" in the forward direction will shift "red" to "blue" in the backward direction, as is apparently the case for the modulator-based device of [37].) Set up this way, the device operation is analogous to looking for a right-circular polarization at the input and setting the device to give a horizontal polarization at the output in the polarization device of Fig. 3.

Now, if we delayed one of the input beams – say the "blue" one – by 180 degrees, the output of Mach-Zehnder MZO would instead appear only on the top waveguide on the right, therefore passing through the frequency shifter and leading to a "red" beam in the output. This could be analogous to changing the input polarization to left-circular and obtaining a vertical polarization at the output in the polarization device of Fig. 3.

This hypothetical device therefore performs the operation

$$\text{"red"} + \text{"blue"} \rightarrow \text{"blue"}$$
$$\text{"red"} - \text{"blue"} \rightarrow \text{"red"}$$

(6)

Though this is an unusual operation for a linear optical device, note that it is linear in the signal field. Note too that the input spectrum "red" + "blue" is orthogonal mathematically to the input spectrum "red" – "blue"; because we can meaningfully define relative phase of two different monochromatic beams, essentially by mathematically comparing the phase of their beat frequency to the phase of a standard drive signal for the frequency shifter, the "+" and "-" signs in Eq. (6) are mathematically meaningful. This device takes orthogonal spectral inputs and maps them to orthogonal spectral outputs, in a two-dimensional spectral space in each case.

This idea of orthogonal spectra is a concept that is not very common in optics because we more typically consider power spectra; because power spectra are always positive, two power spectra can only be simply orthogonal if they do not overlap at all. Here, however, we have two input signals – "red" + "blue" and "red" - "blue" – that have identical power spectra but are nonetheless orthogonal in the sense considered here, and could be used as separate communications channels, for example.

We could imagine extending these concepts to multiple wavelengths; below we discuss in principle how to do so. As an illustration, one concept that then would become possible in principle would be optical spread-spectrum communications. For example, with $N$ different wavelengths, we could construct multiple different spectra, each of which would contain all $N$ wavelengths with equal power, but that were nonetheless orthogonal. (A simple binary approach of inverting the phases in some channels could give $\log_2 N$ such different orthogonal spectra). Such spectra would look the same to a simple spectrometer or to the naked eye, but they could in principle be used simultaneously as separate communications channels, with modulation and detection, using schemes along the lines considered here.

### Universal device

More generally, then, we can expand the idea shown in the simple polarization controller above with other representation converters. Fig. 5 shows an example device configuration in which we first convert from a continuous input field to waveguides using some spatial single mode converters. Then, in this example, we split the polarizations, converting to (twice as many) waveguide modes all in the same polarization. (These two functions could be combined as in the polarization splitting grating couplers discussed above [22].) Next we split each such waveguide mode into separate wavelength components. Finally, we use wavelength converters (frequency shifters) to change each of those components to being at the same wavelength (frequency). Now the input field that was originally a continuous beam with possibly spatially varying polarization content and with multiple frequency components or time-dependence (possibly different for each spatial and polarization component) has been converted into a representation in a set of spatial modes all at the same frequency and polarization. This set of modes is then fed into our device as described above, with the $U^\dagger$ and $V$ blocks representing the self-aligning couplers CI and CO respectively (e.g., in the planar configuration of Fig. 2) and $D_{diag}$ representing the vertical line of modulators SD1, SD2, …, etc. On the right side of the device, we perform the inverse set of representation conversions to that on the left to obtain the final output field.

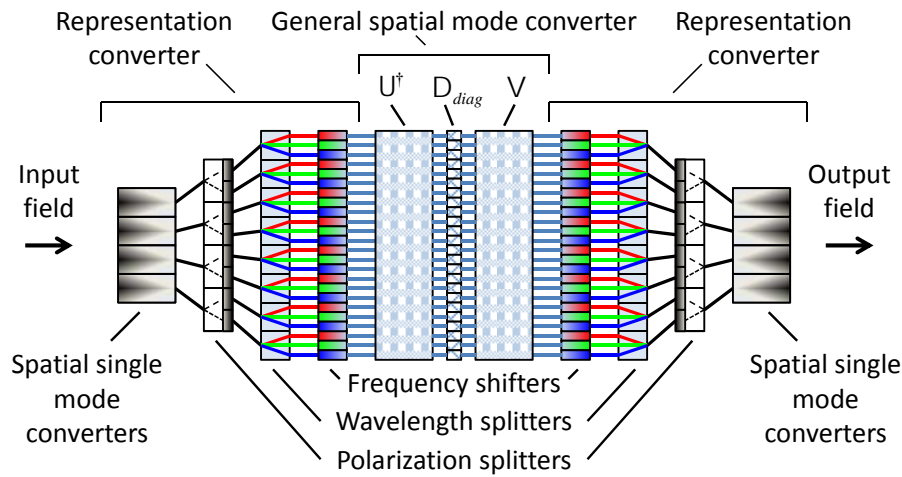

Fig. 5. Example general apparatus for performing arbitrary linear mappings from input fields with spatial, polarization and frequency content to corresponding output fields, illustrated here for 4 spatial modes and 3 different frequency components. Each of the resulting $4 \times 2 \times 3 = 24$ orthogonal channels can be separately modulated using the modulators in the middle column, corresponding to the elements of $D_{diag}$.

Methods for making each of the "representation converter" devices in Fig. 5 are known, at least in principle. Various approaches exist to convert from an input spot or mode to a waveguide mode, including the grating coupler approach (see, e.g., Refs. [2, 14, 15, 31-36]). If we started with a two-dimensional (2D) spatial input field, we could sample it with a 2D array of such spatial single mode converters into optical fibers, and then rearrange the outputs of those fibers into the 1D line of inputs in Fig. 5. Polarization splitters are standard components that can exist in many different forms. Many forms of wavelength splitters, such as gratings, separate different frequencies to different spatial channels.

For a finite input time range or repetition time, we know we can always Fourier-decompose a signal (in a given spatial mode or waveguide) into a set of amplitudes of each of an equally spaced comb of frequencies. We can then, at least in principle (though with greater practical difficulty), convert each frequency component to a standard frequency using frequency shifters [37-40]. As mentioned above, electro-optic frequency shifters, which are conceivable at least for small frequency shifts up to 10's of GHz, could in principle be driven from the beating of the different comb elements, thus retaining well-defined phase relative to the input field. In this way, at least in principle, we can convert an arbitrary Fourier decomposition in different frequency modes emerging from the wavelength splitters into different spatial modes all at the same frequency. (Note, incidentally, that such frequency shifters are linear optical components in that they are linear in the optical field being frequency-shifted; in the case of modulator-based frequency shifters [37], it is largely a matter of taste whether we regard them as being non-linear optical devices in any sense.) Such devices can all, at least in principle, be run backwards at the output to convert frequencies back.

The spatial modes, now all in the same polarization and at the same frequency, pass through the general spatial mode converter (e.g., like Fig. 2). Finally, we pass back through another representation converter to create the output field. In this way, we can in principle perform any linear transformation of the input field, including its spatial, spectral, and polarization forms. As an alternative to frequency shifters, it is possible at least in principle to use time-multiplexing; we discuss this in Appendix D. It is important to note that, to implement an arbitrary linear optical device, it is not sufficient merely to process each frequency or wavelength component on its own without the option of frequency conversion; such a process can implement an arbitrary filter, but it cannot in general perform the linear transformation of one spectrum into another arbitrary spectrum, for example. This point is discussed at greater length in Ref. [15].

he apparatus in Fig. 5 is reminiscent of switching fabrics in optical telecommunications, and this approach can certainly implement the permutations required in such fabrics. The present approach, however, goes well beyond permutations, allowing arbitrary linear combinations of inputs to be mapped to arbitrary linear combinations of outputs, including as other special cases all broadcast and multicast functionalities. Note too that it can in principle perform operations mapping between different kinds of representations, such as converting different orthogonal spatial modes at one frequency at the input into different orthogonal spectra all in the same spatial mode at the output, as well as a many other kinds of linear mappings involving spatial, polarization, and frequency attributes.

### *Devices with forward and backward waves*

So far, we have only considered devices that operate with input waves coming from one side or port and output waves leaving from the other. If the device is to be truly universal, it has to handle waves going in the opposite directions also. If the device function is optically reciprocal, then we can merely run the beams backwards into the device and it will work correctly also in the backward direction. If, however, we want a non-reciprocal function from the device [41] (a Faraday isolator being a simple example), the device as described so far cannot provide such functions. We discuss in Appendix C how further additions of non-reciprocal elements can handle such cases.

### Cloaking

To implement "cloaking" [11-13] in principle, we flow the waveguides (e.g., as optical fibers) connecting any two adjacent vertical blocks of devices Fig. 5 round the volume to be "cloaked" and use the general spatial mode converter to implement the required mapping between input and output fields to emulate free-space propagation through the cloaked volume. Note that, as with all "transmission" cloaks [13], we generally have additional propagation delay that prevents truly perfect cloaking. The overall additional time delay in our universal device is the one sense in which it cannot be made perfect.

### 4.2 Self-configuring operation

So far, for this universal device, we have shown that in principle any such linear transforming device can be made, though we have not explicitly discussed the self-configuration in this general context. The basic principle of self-configuration is not changed for the universal device. We need to take some care when discussing the time-domain behavior, however, when training the output side of the device.

Suppose first that we are operating with the wavelength-splitting version of the universal device, as in Fig. 5. We presume that we work with frequency converters that, when run with waves propagating in the opposite direction, perform the opposite frequency conversion; that is, if when run with a "forward" wave a converter changes the wave frequency from $\omega$ to $\omega + \delta\omega$, then with a wave propagating backwards into it, it will convert from $\omega + \delta\omega$ to $\omega$. The electro-optic frequency converter of [37] can operate in this way, for example. With such a frequency converter, the mapping from spatial to frequency modes and the mapping from frequency to spatial modes are just inverses of one another.

Suppose, then, that we want to train the device to output a pulse $f(t)$ in a particular spatial mode in response to some specific input. Then, in training, we send the same pulse $f(t)$ propagating backwards, i.e., in the phase-conjugated version of the spatial mode. Phase conjugation changes the spatial direction of propagation by changing the sign of the spatial variation of the phase, but it does not time-reverse the pulse envelope (despite the occasional, and somewhat misleading, description of phase conjugation as time-reversal – see [26] for a discussion of this point); the different frequency components in this phase-conjugated pulse have the same relative complex amplitudes at any point in space in both the "forward" and phase-conjugated versions, consistent with the time behavior of the pulse being of the same form. Hence, we need make no change to the frequency splitting and conversion in the apparatus of Fig. 5 to allow it to be self-configuring, as long as the frequency converters operate as discussed here when run backwards.

Self-configuration of the time-multiplexed version of the device is discussed in Appendix D; in that case, a time-reversed pulse should be used during training of the output. For non-reciprocal devices, we have to reverse the circulation direction (e.g., by changing the static magnetic field direction in optical circulators) when training with the backwards beams, as discussed in Appendix C.

## 5. Conclusions

In conclusion, we have shown that there is at least one constructive method to design an arbitrary linear optical component capable in principle of any spatial, polarization, and spectral linear mapping, in any combination. This method can also be self-configuring, extending the concepts of the self-aligned universal wave coupler [21]. Only local feedback loops, optimizing one parameter at a time, are required. This feedback-based operation avoids the necessity of setting calculated analog values with interferometric precision in collections of optical components. This approach can also allow simultaneous and separately modulated conversions from multiple orthogonal inputs to corresponding orthogonal outputs. Though discussed here in the language and technology of optics, the method can be extended to other linear wave problems generally, including radio-frequency electromagnetics, acoustics, and quantum mechanical waves and superpositions. Versions for certain specific optical uses, such as arbitrary polarization and spatial mode conversions and modulations, appear practical with current planar optical technology.

*Acknowledgements*

This project was supported by funds from Duke University under an award from DARPA InPho program, and by Multidisciplinary University Research Initiative grants (Air Force Office of Scientific Research, FA9550-10-1-0264 and FA9550-09-0704). I am pleased to acknowledge discussions with Mohammad Mirhosseini.

## Appendix A. Progressive calculation of reflectivities and phase shifts

Though the device can operate in a self-configuring mode, we can also formally calculate what the reflectivities and phases need to be in all of the beamsplitter blocks. Fig. 6 shows one unitary transformer (here for $\mathsf{U}$) with the reflectivities and phase shifts labeled, analogous to coupler CI in Fig. 1. (Detectors are omitted here.)

The reflectors and phase shifters in Fig. 6 (and in Fig. 1) are shown as rectangles only in the middle of the beamsplitter blocks, but it is understood that they act on the entire beam passing through each block. A completely arbitrary unitary transformer would require the phases shifters at the right in the dashed rectangles so as to set the overall phases of the outputs on the right, and we will use these in our algebra here, though we do not need these in the architecture of Fig. 1 because the singular value modulators SD1 – SD4 can set any specific phase required between the beamsplitter blocks for $\mathsf{U}$ and $\vee$.

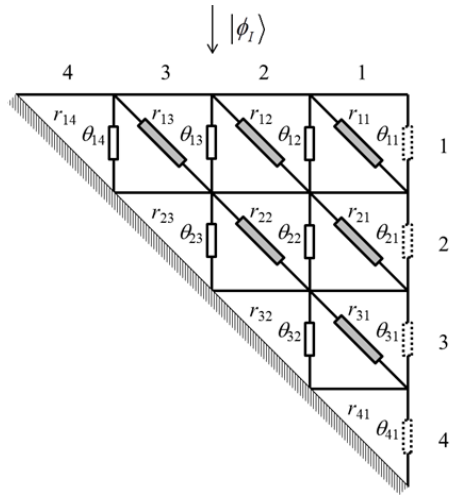

Fig. 6. The mode transformer for the operator $\mathsf{U}$ for $M = 4$ with the reflectivities and phase shifts labeled for each beamsplitter block. The diagonal mirror has 100% reflectivity.

To discuss the phases involved in the beamsplitter, we need some formal definitions. Fig. 7 shows a (lossless) beamsplitter (without any additional phase shifter). We can define complex field transmission factors $t^{(TB)}$ from top to bottom and $t^{(LR)}$ from left to right, and similarly define field reflection factors $r^{(TR)}$ and $r^{(LB)}$. These complex factors include the phase shifts between the respective inputs and outputs as their arguments: for example, the phase delay between top and bottom is $\theta^{(TB)}$ in the expression

$$t^{(TB)} = \left| t^{(TB)} \right| \exp\left( i\theta^{(TB)} \right) \qquad (7)$$

and similarly for the other transmission and reflections.

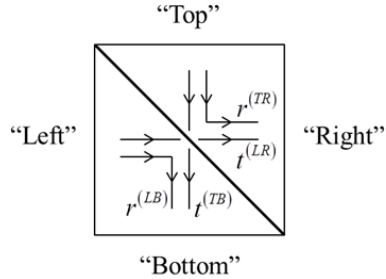

Fig. 7. Beamsplitter with definitions of field reflection and transmission factors and nominal labels of the beamsplitter ports as top, bottom, left and right.

Because the beamsplitter is lossless [42]

$$\left| t^{(TB)} \right|^2 = 1 - \left| r^{(TR)} \right|^2 = \left| t^{(LR)} \right|^2 = 1 - \left| r^{(LB)} \right|^2 \qquad (8)$$

and, obviously from Eq. (8), $\left| r^{(TR)} \right|^2 = \left| r^{(LB)} \right|^2$. Also,

$$\theta^{(TR)} + \theta^{(LB)} - \theta^{(TB)} - \theta^{(LR)} = \pm \pi \qquad (9)$$

(at least within some additive phases in units of $2\pi$, which we neglect for simplicity in the algebra).

We will formally write any of our input basis functions $\left| \phi_{DIm} \right\rangle$ as a linear combination of the "modes" (rectangular functions) corresponding to the inputs to the individual columns

$$\left| \phi_{DIm} \right\rangle = \sum_{n=1}^{M} a_{mn} \left| \phi_{1n} \right\rangle \qquad (10)$$

where by $\left| \phi_{1n} \right\rangle$ we mean the (input) mode (rectangular function) incident on the top row in the $n$th column.

As discussed in the main text, the idea of this unitary transformer is that, if we illuminate from the top with the function $\left| \phi_{DI1} \right\rangle$, all the power will come out of port 1 at the right. Similarly, illuminating with function $\left| \phi_{D I 2} \right\rangle$ will lead to all the power coming out of port 2 at the right, and so on. To understand how to set the reflectivities $r$ and phase shifts $\theta$ in the top row mathematically, we imagine for the moment that we are running the device backwards, shining a beam into port 1 on the right and looking at the beams coming out of the ports at the top. We presume that we are dealing only with reciprocal optics in our beam splitters and phase shifters so that the phase delays and the magnitudes of the reflectivities are the same forwards and backwards. The output amplitudes that we want our device to generate at the top in this backwards case should therefore be the complex conjugates $a_{1n}^*$ of the amplitudes in Eq. (10); if we generate some phase delays in running the device backwards, then we should have corresponding phase leads in the input beams when running the device forwards so all the beams add up with the correct phase at output 1 on the right.

Hence, for the top right block in Fig. 6, we should choose

$$r_{11}^{(TR)} \exp\left(i\theta_{11}\right) = a_{11}^* \tag{11}$$

In operation, when we choose the magnitude of a given $r^{(TR)}$, for example by setting phase delay in a Mach-Zehnder interferometer implementation of a variable beam splitter, the phase $\theta^{(TR)}$ associated with $r^{(TR)}$ will also be set as a result and we will know what it is. (Note in our mathematics here we are allowing for possible changes in phase associated with changes in reflectivity, though in the self-configuring versions of the device discussed in the main text, we prefer to work with components that do not change phase as they change reflectivity because it makes the feedback loops simpler.) We will then choose the phase shifter phase delay (e.g., the $\theta_{11}$ in Eq. (11)) so as to satisfy the necessary overall design requirement on phase, as in Eq. (11) here.

Now knowing $r_{11}^{(TR)}$ (and hence, from Eq. (8), also $t_{11}^{(LR)}$) and $\theta_{11}$, we can proceed to the next block in this first row. The field that will emerge from top in the second column is

$$t_{11}^{(LR)} r_{12}^{(TR)} \exp\left[i\left(\theta_{11} + \theta_{12}\right)\right] = a_{12}^* \tag{12}$$

so we should choose

$$r_{12}^{(TR)} \exp\left(i\theta_{12}\right) = a_{12}^* \exp\left(-i\theta_{11}\right) / t_{11}^{(LR)} \tag{13}$$

We can continue progressively along the top row, with the reflectivity and phase in the *n*th column being chosen to satisfy

$$r_{1n}^{(TR)} \exp\left(i\theta_{1n}\right) = a_{1n}^* \exp\left(-i \sum_{p=1}^{n-1} \theta_{1p}\right) / \prod_{q=1}^{n-1} t_{1q}^{(LR)} \tag{14}$$

where we understand that when $n = 1$ the summation term will be 0 and the product term will be 1. (Note that the magnitude of the last reflectivity, $\left| r_{1M}^{(TR)} \right|$, will always be 1, which is ultimately guaranteed by the lossless nature of this set of beamsplitters and the consequent unitarity of the operators.)

Now we consider what happens when we shine the second basis function $\left|\phi_{DI2}\right\rangle$ into the top of the set of beamsplitters. First we need to set up some notation. For a field arriving at the top of the $u$th row of beamsplitter blocks, we can choose to write

$$\left|\phi^{(u)}\right\rangle = \sum_{j=1}^{M-u+1} a_j^{(u)} \left|\phi_{uj}\right\rangle \tag{15}$$

where, in an extension from the kind of notation used in Eq. (10), by $\left|\phi_{uj}\right\rangle$ we mean the (input) rectangular "mode" incident on the **$u$**th row in the **$j$**th column. Given that we know all the reflectivities (and hence trasmissivities) and phases of the first row of beamsplitter blocks, given some field $\left|\phi^{(1)}\right\rangle$ incident on the top row, we can deduce what field $\left|\phi^{(2)}\right\rangle$ will arrive at the top of the second row. We can formally write this linear relation in terms of a matrix $\mathrm{C}^{(1)}$

$$\left|\phi^{(2)}\right\rangle = \mathrm{C}^{(1)}\left|\phi^{(1)}\right\rangle \tag{16}$$

where $\mathrm{C}^{(1)}$ is the first of a family of $(M-u)\times(M-u+1)$ matrices

$$\mathrm{C}^{(u)} = \begin{bmatrix} t_{u1}^{(TB)} & c_{12}^{(u)} & c_{13}^{(u)} & \cdots & c_{1(M-u)}^{(u)} & c_{1(M-u+1)}^{(u)} \\ 0 & t_{u2}^{(TB)} & c_{23}^{(u)} & & & \\ \vdots & 0 & t_{u3}^{(TB)} & \ddots & \vdots & \vdots \\ & \vdots & \ddots & \ddots & & \\ 0 & 0 & 0 & \cdots & t_{u(M-u)}^{(TB)} & c_{(M-u)(M-u+1)}^{(u)} \end{bmatrix} \tag{17}$$

where $c_{sj}^{(u)}$ is the "complex fraction" (i.e., the multiplier) of the field incident on column **$j$** of row **$u$** that contributes to the field incident on the top of column **$s$** of row $u+1$. For the diagonal elements,

$$c_{ss}^{(u)} = t_{us}^{(TB)} \tag{18}$$

For the elements to the right of the diagonal,

$$c_{sj}^{(u)} = r_{uj}^{(TR)} r_{us}^{(LB)} \left[\prod_{p=s+1}^{j-1} t_{up}^{(LR)}\right] \exp\left[i \sum_{p=s+1}^{j} \theta_{up}\right] \tag{19}$$

This element is the product of (i) the field reflectivity $r_{uj}^{(TR)}$ of the "sideways" reflecting beamsplitter in block $uj$ that reflects into row $u$, (ii) the field reflectivity $r_{us}^{(LB)}$ in the "downwards reflecting" beamsplitter in block $us$ that reflects down into row $u+1$, (iii) the product of all the "sideways" transmissions in all the intervening blocks, and (iv) the phase factors from all of the phase shifters encountered on this path.

So, given that we have calculated all the reflectivities and phases for the first row, we can now calculate $C^{(1)}$, and hence when we shine the second basis function $|\phi_{DI2}\rangle$ onto the top of the whole device, we will obtain a field

$$\left|\phi_{DI2}^{(2)}\right\rangle \equiv \sum_{j=1}^{M-1} a_{2j}^{(2)} |\phi_{2j}\rangle = C^{(1)} |\phi_{DI2}\rangle \tag{20}$$

at the top of the second row.

Now to calculate the settings of the reflection and phase factors for the second row, we proceed in a similar fashion to that used for the first row, but with input amplitudes on the top of the $n$th column of the second row of $a_{2n}^{(2)}$ instead of the amplitudes $a_{1n}$ we used in calculating the first row reflection and phase factors.

For the third row, having calculated all the reflections and phases in the second row, we can calculate the matrix $C^{(2)}$ and hence calculate amplitudes $a_{3n}^{(3)}$ that will appear at the top of the third row when we illuminate the top of the device with the third basis function $|\phi_{DI3}\rangle$

$$\left|\phi_{DI3}^{(3)}\right\rangle \equiv \sum_{j=1}^{M-2} a_{3j}^{(3)} |\phi_{3j}\rangle = C^{(2)}C^{(1)} |\phi_{DI3}\rangle \tag{21}$$

We proceed similarly to calculate progressively all subsequent rows, thereby completing the design mathematically.

Note that shining the second basis input $|\phi_{DI2}\rangle$ on the top of the structure produces no output from port 1 on the right. The unitarity of the overall operation means that orthogonal inputs always give orthogonal outputs (unitarity preserves all inner products). Because $|\phi_{DI2}\rangle$ is orthogonal to $|\phi_{DI1}\rangle$, then their outputs must also be orthogonal. Since the output with $|\phi_{DI1}\rangle$ is solely from the top port, $|\phi_{DI2}\rangle$ can therefore have no component emerging from the top port. Similar behavior follows for all subsequent orthogonal inputs, each of which leads only to output from one (different) port at the right of the structure.

To calculate the reflections and phases in the device implementing the unitary transformation $V$, for which we want output functions

$$\left|\phi_{DOm}\right\rangle = \sum_{n=1}^{M} b_{mn} |\beta_{1n}\rangle \tag{22}$$

where by $|\beta_{uj}\rangle$ we mean the (output) mode leaving the top of the $u$th row in the $j$th column, we can proceed similarly. Here, when we shine light into a port on the left of the output coupler structure (as in CO in Fig. 1 of the main text), we want to create the actual output fields for a given output basis function, so we do not take the complex conjugates of the amplitudes $b_{mn}$ for our calculations. That is, where we have $a_{mn}^*$ in Eqs. (11) - (14), we will use $b_{mn}$ in the analogous equations for $\vee$.

## Appendix B. Implementation with Mach-Zehnder interferometers

The Mach-Zehnder waveguide modulator [43] configuration used in the main text as in Fig. 2 implements the necessary control of reflectivity and phase using two phase shifters within the modulator. Fig. 8 shows the modulator configuration in detail. The phase shifting could be accomplished with electrooptic materials with voltages applied through electrodes or with thermal devices, which here for simplicity of description we take to have phase shift also set by some voltage. (For such thermal phase shifters, negative voltages would not, however, give negative phase shifts, so in that case, we can imagine the voltages we discuss here to be in addition to some positive bias so that all actual voltages are positive in the thermal case).

Nominally defining the phase delays in the phase shifters as being between points C and F (D and G) for the upper (lower) phase shifter, the average voltage controls the common-mode phase shift $\theta_{av}$ and the difference between the voltages controls the differential phase shift $\Delta\theta$. The device is presumed perfectly symmetric; in a real device we might add one or more control phase shifting electrodes inside the beamsplitter sections to achieve symmetric behavior in practice. Here we formally analyze the Mach-Zehnder interferometers, showing how to relate their behavior and settings to those of the "conventional" beam splitters and phase shifters of Fig. 1 and the discussion of Appendix A on the required values in an actual design.

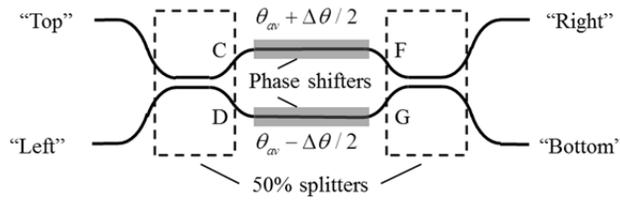

Fig. 8. Symmetric Mach-Zehnder waveguide modulator configuration with 50% ("3 dB") splitters notionally implemented here with coupled waveguides and two arms each with a phase shifting element. The grey rectangles represent the phase shifting control elements (e.g., electrodes). The labeling of the ports corresponds with the notation used in Fig. 7.

In a symmetric Mach-Zehnder device as in Fig. 8, the 50% splitters are each identical symmetrical loss-less beam splitters. Reflection within these 50% splitters corresponds to the paths "Top" – C, "Left – D", F – "Right", and G – "Bottom". The phase delays associated with these reflections, $\theta_{TC}$, $\theta_{LD}$, $\theta_{FR}$, and $\theta_{GB}$, respectively are all equal, i.e.,

$$\theta_{refl} = \theta_{TC} = \theta_{RD} = \theta_{FL} = \theta_{GB} \tag{23}$$

Similarly for the transmission phases, with obvious notation,

$$\theta_{trans} = \theta_{TD} = \theta_{LC} = \theta_{FB} = \theta_{GR} \tag{24}$$

Similarly, the magnitudes of the various transmissions and reflections through these 50% splitters are all equal at a value $1/\sqrt{2}$ (which leads to the 50% power splitting). There may be an additional fixed phase delay $\theta_{ex}$ associated with any other waveguide propagations not accounted for in phase delays in the 50% splitters and the phase shifters.

Adding the fields on the two "transmission" paths through the different 50% splitters and phase shifters, the overall complex field transmissions $t^{(TB)}$ and $t^{(LR)}$ are both therefore given by

$$t^{(TB)} = t^{(LR)} = t \exp\left(i\theta_S\right)\exp\left(i\theta_{av}\right) \tag{25}$$

where

$$t = \cos\left(\Delta\theta / 2\right) \tag{26}$$

and the background "static" phase $\theta_S$ is the sum

$$\theta_S = \theta_{ex} + \theta_{trans} + \theta_{refl} \tag{27}$$

Before adding up the phases for the reflection paths, we note from Eq. (9) above, with Eqs. (23) and (24), that we can write

$$\theta_{trans} = \theta_{refl} \pm \pi / 2 \tag{28}$$

Whether we use the "+" or the "-" here depends on the detailed design of the 50% splitters. (It is also possible in principle that there are additional amounts of phase in units of $\pi$ that could be added to the right of Eq. (28), but we neglect those for simplicity.) Adding the fields on the two "reflection" paths, we obtain

$$r^{(TR)} = -r^{(LB)} = \mp r \exp\left(i\theta_S\right)\exp\left(i\theta_{av}\right) \tag{29}$$

where

$$r = \sin\left(\Delta\theta / 2\right) \tag{30}$$

In formally designing using this kind of dual phase-shifter Mach-Zehnder device, we can drop the additional phase factors of the form $\exp\left(i\theta_{up}\right)$ as in Eqs. (11) - (14) and (19), because all the necessary phase factors are included in the field reflection and transmission coefficients $r^{(TR)}$, $r^{(LB)}$, $t^{(TB)}$ and $t^{(LR)}$. We use the choice of $\Delta\theta$ to set the magnitude of $r^{(TR)}$ and the choice of $\theta_{av}$ sets its phase, with the magnitudes and phases of $r^{(LB)}$, $t^{(TB)}$ and $t^{(LR)}$ being therefore set also.

When used as an amplitude modulator as part of implementing the singular values $s_{Dm}$ in an architecture such as that of Fig. 2 of the main text, the power out of the "bottom" port will be dumped.

## Appendix C. Non-Reciprocal devices

To handle non-reciprocal optical elements in this approach, or any element where we want separate control of forward and backward waves in the ports of the device, we can in principle add forward/backward splitters to the left and right sides of the apparatus of Fig. 5 as shown in Fig. 9; the example configuration in Fig. 9 shows a general 4-port optical device with input and output modes in all 4 ports.

This example approach is based on the use of 3-port optical circulators [44-45] to separate forward and backward waves. Backward waves coming into the right of the structure are separated from the forward waves and fed as additional inputs into the left of the general spatial mode converter in the middle. Two of the four outputs from the general spatial mode converter are fed to the optical circulators on the left to give the backward propagating output beams on the left.

The addition of such circulator devices, which are non-reciprocal by definition, allows the whole optical arrangement to be non-reciprocal if required, while leaving the core general spatial mode converter itself as a reciprocal device that always runs only from front to back (left to right). We could add circulator optics to the apparatus of Fig. 3, for example, by putting the circulators between the polarization and wavelength splitters in half of the channels on each side, in a fashion similar to that of Fig. 9.

For self-configuration using the non-reciprocal device approach of Fig. 9, during training for setting the output $\vee$ coupler with the reversed versions of the desired output beams, we need to reverse the sense of the circulators; i.e., the rotation arrows should be flipped from clockwise to anticlockwise at the input and from anticlockwise to clockwise at the output. Such a change might be achieved by changing the direction of the static magnetic fields in circulators based on Faraday isolation.

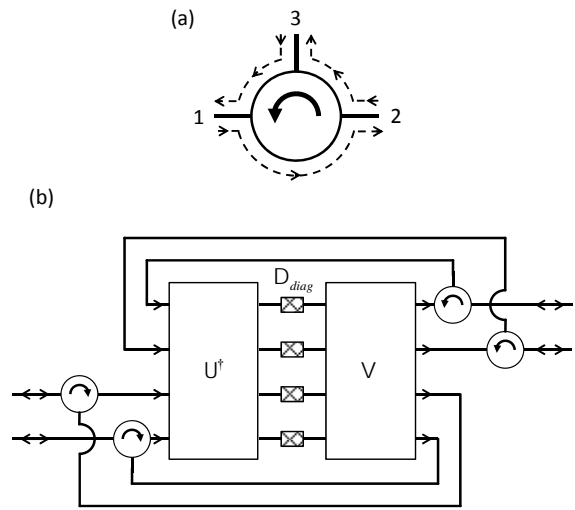

Fig. 9. Use of optical circulators with forward and backward modes. (a) Schematic of a 3-port optical circulator. The dashed lines show the effective paths of waves in different directions between the three ports. (b) Universal 4-port "two-way", potentially non-reciprocal device, with input and output beams in each of two paths at both the left and right of the device. The central "$\cup^{\dagger}$", "$D_{diag}$", and "$\vee$" units form a general spatial mode converter as in Figs. 1, 2, and 4.

## Appendix D. Time-Multiplexing Representation converters

As an alternative to the frequency splitting and frequency conversion of Fig. 5, in principle we could split an input pulse into different time windows, then pass each of those through the general spatial mode converter. Idealized time delay units for implementing a time (rather than frequency) version of the approach are shown in Fig. 10. At the input side, the paths connected to points 2 and 1 have additional propagation delays compared to the path connected to point 3 of $\Delta t$ and $2\Delta t$, respectively.

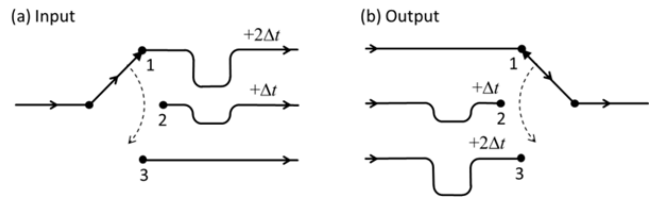

Fig. 10. Illustration of an idealized time delay unit. The switches rotates through positions 1, 2, and 3, with a dwell time of $\Delta t$ at each position, taking a total time of $3\Delta t$ to cycle through all 3 positions before returning to position 1. (a) Switch used at input side. (b) Switch used at output side.

Thus the signals from three successive time windows of duration $\Delta t$ appear simultaneously at the three outputs on the right in Fig. 10(a), allowing them then to be fed into the general spatial mode converter (or into the next stage of the preparatory representation conversion stages). A similar apparatus can be used at the output, but operated with the delays reversed to reconstruct a signal segment of duration $3\Delta t$ at the final output, with each $\Delta t$ time slot in that signal being an arbitrary linear combination of 3 incident $\Delta t$ time slots. See [46] for a summary of time multiplexing schemes and [47] for a recent example, though many such schemes also convert frequencies, which is not desirable here.

If we are operating using the time-domain rather than frequency-domain devices, i.e., using units as in Fig. 10 rather than the wavelength splitters and converters of Fig. 5, and we want to train the device to output a pulse of temporal form $f(t)$ for a given input, then, at least if using the time-delay units of Fig. 10, we would need to train with a time-reversed pulse, i.e., of form $f(-t)$ running in each spatial mode back into the device; otherwise we do not get the desired relative delays of each segment of the pulse so that they are all lined up in time within the central general spatial mode converter.